\documentclass{article}
\usepackage[utf8]{inputenc}
\usepackage{amsmath}
\usepackage{amsfonts}
\usepackage{braket}
\usepackage{graphicx}
\usepackage{physics}
\usepackage{xcolor}
\usepackage{multirow}
\usepackage{adjustbox}
\usepackage[a4paper, total={6in, 8in}]{geometry}
\usepackage[labelformat=empty]{caption}
\title{Non-separable Optical Beam Shifts and Emergence of Position-position Classical entanglement}
\author{Niladri Modak$^{1,*}$, S Ashutosh$^{1}$, Shyamal Guchhait$^{1}$, Sayantan Das$^{1}$, Alok Kumar Pan$^{2}$, \\Nirmalya Ghosh$^{1,3,+}$\\
$^{1}$\textit{\small Department of Physical Sciences},\\ \textit{\small Indian Institute of Science Education and Research Kolkata,}\\ \textit{\small Mohanpur, India- 741246}\\
$^{2}$\textit{\small Physics Department, National Institute of Technology Patna, Patna},\\\textit{\small India- 800005}\\
$^{3}$\textit{\small Centre of Excellence in Space Sciences India},\\ \textit{\small Indian Institute of Science Education and Research Kolkata,}\\ \textit{\small Mohanpur, India- 741246}}

\date{\footnotesize$^{*}$nm16ip018@iiserkol.ac.in, $^{+}$nghosh@iiserkol.ac.in}

\begin{document}

\maketitle

\begin{abstract}
\noindent
 Under the introduction of any interface in its trajectory, an optical beam experiences polarization-dependent deflections in the longitudinal and transverse directions with respect to the plane of incidence. The physics of such optical beam shifts is connected to profound universal wave phenomena governed by the fine interference effects of wave packets and has opened up avenues towards metrological applications. Here, we reveal the inherent non-separability of the longitudinal and transverse beam shifts by considering a rather simple case of a partially reflecting Gaussian laser beam from a dielectric interface. This non-separability appears substantially in some particular regions in the corresponding parameter space. We further show that such non-separability manifests as a position-position classically entangled state of light. The tunability of the related experimental parameters offers control over the degree of entanglement. Uncovering of the inherent non-separability of the two types of beam shifts is expected to enrich the physical origin of this fundamental effect, impact the understanding of numerous analogous effects, and might find useful applications by exploiting the position-position-polarization classical entanglement in a fundamental Gaussian beam.

\end{abstract}
\section*{Main Text}
The spatial degree of freedom (DoF) of any optical beam is known to be classically entangled with its polarization \cite{simon2010nonquantum}. However, as far as elementary beams are concerned, the two DoFs of a sufficiently broad beam can be safely treated in a separable fashion \cite{simon2010nonquantum}. Such a beam, when partially or total internally reflected from an interface, experiences a weak non-separability between polarization and spatial DoFs \cite{bliokh2013goos,toppel2013goos}. Spatially inhomogeneous polarization distribution after such interactions is at the heart of such non-separability, leading to polarization-dependent shifts of the centroid of the beam  \cite{bliokh2013goos,toppel2013goos}. These beam shifts may occur either in the plane of incidence (longitudinal), known as Goos–H\"anchen (GH) shift, and /or in its perpendicular (transverse) direction, the so-called Imbert–Federov (IF) shift or spin Hall effect of light \cite{bliokh2013goos,toppel2013goos}.The GH shift appears due to the dispersion of Fresnel reflection and transmission coefficients, which has pure dynamical origin. On the other hand, IF shift originates from the spin-orbit interaction of light and appear due to the spatial or momentum gradient of geometric phase \cite{bliokh2013goos}. The physics of these optical beam shifts is not just mere corrections to Snell’s law for plane waves but are deeply connected with a number of non-trivial 
wave phenomena originating from the interference of either classical electromagnetic waves or quantum matter waves. 
Weak measurements and weak value amplification \cite{toppel2013goos}, superoscillations \cite{berry2019roadmap}, Wigner time delay \cite{chauvat2000direct}, super and sub-luminal propagation of wave packets \cite{asano2016anomalous}, spin-orbit interaction of light \cite{bliokh2015spin}, PT symmetry \cite{gotte2014eigenpolarizations}, and classical entanglement between spatial mode and polarization degree of freedom \cite{toppel2013goos} are 
a plethora of intriguing wave phenomena that are encoded in the physics of optical beam shifts. The prospect of gaining fundamental insights into these wave phenomena in relatively clean and simple optical systems and the possibility of extrapolating results to a range of physical systems has triggered enormous interest in studying these optical beam shifts. 
These shifts have been well demonstrated in a wide variety of systems ranging from dielectrics \cite{hosten2008observation,goswami2014simultaneous,pal2019experimental}, metallic surfaces to meta-materials and multilayered structures \cite{ye2019goos,grosche2015goos,kong2019goos}. Due to the fundamentally different origins of the longitudinal GH and transverse IF shifts, these have been treated in a completely separable manner in all the previous reports \cite{toppel2013goos,goswami2014simultaneous,ye2019goos,grosche2015goos,kong2019goos,gotte2012generalized,gotte2014eigenpolarizations}. Here, we reveal inherent non-separability of the longitudinal and the transverse beam shifts by considering a rather simple case of a partially reflecting fundamental Gaussian beam. We observe that such non-separability appears substantially in some particular region of the corresponding parameter space and can be estimated through some typical quantities which seem to be missing in the investigations so far. Importantly, we go on to demonstrate that this non-separability of the longitudinal and the transverse DoFs of the Gaussian beam leads to a position-position ``classically entangled state" \cite{karimi2015classical,shen2022nonseparable,spreeuw1998classical}.

\par
The typical property of entanglement, be it quantum or classical, is ascribed to the non-separability between two or more sub-systems constituting the whole system \cite{karimi2015classical,shen2022nonseparable,forbes2019classically,schrodinger1935discussion,einstein1935can,horodecki2009quantum}. Although non-locality is an exclusive feature of quantum entanglement \cite{schrodinger1935discussion,einstein1935can,horodecki2009quantum,groblacher2007experimental}, non-separability between more than one DoFs of a single system is well observed in classical optics and termed ``classical entanglement" \cite{shen2022nonseparable,forbes2019classically,karimi2010spin}. In recent times, it has attracted a lot of attention in terms of potential applications and fundamental insights \cite{marrucci2006optical,ndagano2017characterizing,otte2018entanglement,gailele2018free,liu2020multidimensional,hu2021free,zdagkas2020space,devlin2017arbitrary,shen2021creation}. Classical entanglement has been demonstrated in the context of various DoFs of light, e.g., spatial mode, polarization, trajectory, time, frequency, of a single optical system \cite{shen2022nonseparable,toppel2014classical} through different optical effects such as vector vortex beams \cite{marrucci2006optical,ndagano2017characterizing,otte2018entanglement}, polarization dependent beam shifts \cite{toppel2013goos}, spatio-temporal pulses etc \cite{zdagkas2020space}. Most of these demonstrations require precise engineering, such as spatial structuring of vector vortex beams  \cite{marrucci2006optical,ndagano2017characterizing,otte2018entanglement,bliokh2009goos}, structured materials or judicious manipulation of light matter interactions \cite{shen2022nonseparable,zdagkas2020space,shen2021creation,puentes2004optical}. In contrast, the position-position classical entanglement naturally arises here from the non-separability of the longitudinal and transverse beam shifts through one of the most trivial light-matter interactions, the partial reflection of a fundamental Gaussian beam. Regulated control over the degree of entanglement is achieved by tuning the experimental parameters. The emergence of such entanglement is therefore quite ubiquitous and can be observed in a broad range of light-matter interactions \cite{born2013principles}.
\begin{figure}[h!]
    \centering
    \includegraphics[width=0.8\textwidth]{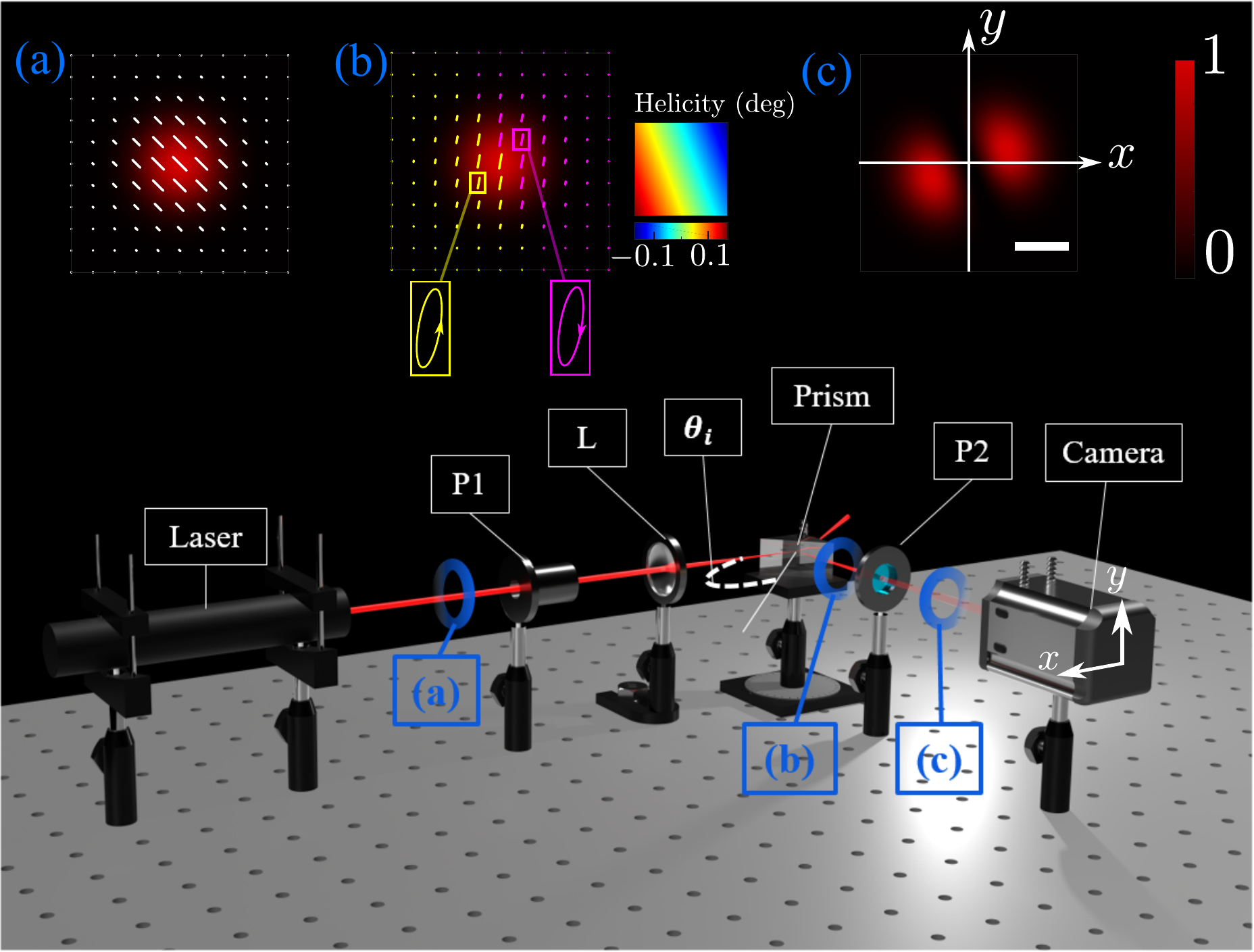}
    \caption{\textbf{Figure. 1: Schematic illustration of the origin of the non-separability of longitudinal and transverse beam shifts and subsequent manifestation of position-position classical entanglement in a Gaussian beam partially reflecting from an air-glass interface}. Schematic of the experimental setup. The fundamental Gaussian mode of a He-Ne Laser is passed through a Glan-Thompson polarizer P1, a 75 mm focal length biconvex lens L and is externally reflected from a $45^{o}$-$90^{o}$–$45^{o}$ BK7 Prism (refractive index 1.516). Angle of incidence is $\theta_i$. P1 determines the input polarization state. P2 is post-selecting sheet polarizer, mounted on a precision rotation mount. The intensity of the beam is measured in the CCD. Insets show the theoretically simulated (using Eq.\eqref{eq4}) beam structure (in local $x-y$ coordinate system) and corresponding spatial distribution of polarization for $\theta_i=50^{o}$, input state $-45^{o}$: (a) the input Gaussian beam having uniform polarization distribution (white lines); (b) partially reflected beam with weak inhomogeneous polarization distribution \cite{gotte2012generalized} and the corresponding polarization helicity \cite{gupta2015wave} plot, magenta ellipse:  negative helicity, yellow ellipse: positive helicity, two representative ellipses are scaled for better visual understandings; (c) non-separable longitudinal and transverse shift manifesting as position-position classical entanglement in the post-selected beam \cite{toppel2014classical}. White scale bar is $250\mu m$.}
    \label{fig1}
\end{figure}

\par 
We consider a polarized Gaussian optical beam undergoing partial reflection from an air-glass interface at an angle of incidence $\theta_i$ (Fig.\ref{fig1}). The transverse profile of the beam can be expressed as a function of local Cartesian coordinate $x-y$, where $x$ is the coordinate in the plane of incidence, while $y$ is perpendicular to this plane. To study the non-separability of GH and IF shift through such a 2D wave function, we start with a general function $F(x,y)$. These variables $x$, and $y$  can be thought of as two DoFs representing two subsystems of a composite system represented by $F(x,y)$. The corresponding subsystems are separable if $F(x,y)$ can be factorized into two independent functions, say, $F(x,y)\equiv f(x) g(y)$. On the other hand, the non-factorizability of $F(x,y)$ implies non-separability of the subsystems \cite{forbes2019classically}. Now, we define a quantity $\rho$ as a measure of the factorizability of such functions. 
\begin{equation}
   \rho = \frac{\langle xy \rangle -\langle x \rangle \langle y \rangle}{\sqrt{\langle x^2 \rangle - \langle x \rangle^2}\sqrt{\langle y^2\rangle - \langle y \rangle^2}}
   \label{eq1}
\end{equation}
Note that the mathematical expression of $\rho$ is identical to that of the Pearson correlation coefficient widely used in statistical data analysis \cite{benesty2009pearson,jebarathinam2020pearson}. Here, $\langle (\ldots) \rangle$ ($=\frac{\iint F^* (\ldots) F\ dx\ dy}{\iint F^* F\ dx\ dy}$) is the average value of $(\ldots)$ treating $F(x,y)$ as a distribution. Clearly, $|\rho|=0$($1$) indicates fully (non-)separable $F(x,y)$. As $\rho$ qualifies as a measure of non-separability, we shall use this regarding beam shifts.

\par
Now, coming back to the context of optical beam shift, consider the polarization of the incident beam (input polarization) is $[E_x,E_y]^T$ (written in $x-y$ basis). The reflected electric field, $\vec{E}_{ref}$ comprises of an inhomogeneous polarization distribution in its transverse plane \cite{pal2019experimental,aiello2008role} as given below.
\begin{equation}
  \vec{E}_{ref}\sim G(x,y)
  \begin{pmatrix}
  r_{p}(1-\frac{ix}{z_{o}+iz}\frac{\partial lnr_{p}}{\partial \theta_i})E_x+ \frac{iy}{z_{o}+iz}(r_{p} + r_{s})\cot{\theta_i}E_y\\\frac{-iy}{z_{o}+iz}(r_{p} + r_{s})\cot{\theta_i}E_x+ r_{s}(1-\frac{ix}{z_{o}+iz}\frac{\partial lnr_{s}}{\partial \theta_i})E_y
  \end{pmatrix}
  \label{eq2}
\end{equation}
Here, $G(x,y)$ is the Gaussian profile of the incident field given by $G(x,y)\sim \exp{k(iz-\frac{x^2+y^2}{z_0+iz})}$, $r_p$ and $r_s$ are the Fresnel reflection coefficients, $z_{o}$ denotes the Rayleigh range and $z$ is the propagation distance after reflection, $k$ is the wave number of the incident electric field \cite{born2013principles,aiello2008role}. As evident from Eq.\eqref{eq2}, both the elements of the Jones vector \cite{born2013principles} include both the spatial DoFs $x$, and $y$. Hence, the apparently separable (in $x$, $y$, and polarization) incident Gaussian beam \cite{simon2010nonquantum} (Fig.\ref{fig1} inset (a)), after reflection, includes three non-separable DoFs, i.e., longitudinal position $x$, transverse position $y$, and polarization (Fig.\ref{fig1} inset (b)). When $\vec{E}_{ref}$ is post-selected (Fig.\ref{fig1} inset (c)), i.e., projected onto any arbitrary polarization state, the final state becomes $\ket{\psi_f(x,y)} \sim G(x,y) F(x,y)$. Here, $F(x,y)$ acts as a spatial response function \cite{asano2016anomalous,pal2019experimental} and takes the form, say,
\begin{equation}
F(x,y)\sim a+bx+cy
\label{eq3}
\end{equation}
where the coefficients $a$, $b$, $c$ are dependent on the experimental parameters $r_p$, $r_s$, $\theta_i$, input and post-selected polarization state \cite{aiello2008role}. Eq.\eqref{eq3} suggests that after reflection and post-selection, the $x$ and $y$ profiles of the beam undergo certain modifications governed by $F(x,y)$, formally named as GH and IF shift respectively \cite{bliokh2013goos,aiello2008role}. Conventionally, beam shifts are measured as expectation values of the corresponding shift operators \cite{toppel2013goos}, i.e., the input state and post-selection are chosen to be same \cite{toppel2013goos,gotte2012generalized,gotte2014eigenpolarizations}. However, for partial reflection, such beam shifts are very small \cite{bliokh2013goos} and ``weak value amplification" (WVA) \cite{aharonov1988result,duck1989sense} protocols are applied for their detection \cite{hosten2008observation,goswami2014simultaneous,pal2019experimental}. In such a case, the input state is chosen to be a superposition of the eigenpolarizations of the corresponding shift operators, and the post-selection is near orthogonal to the input one \cite{hosten2008observation,goswami2014simultaneous,pal2019experimental}. Also, shifts can be observed, in general, with any arbitrary input state and post-selection \cite{gotte2012generalized}. Note that, in all the aforementioned cases, the structure (Eq.\eqref{eq3}) of the spatial response function $F(x,y)$ is generic and in principle, non-separable irrespective of the input and post-selected polarization state. However, the coefficients $a,b,c$ would be different depending on the specific light-matter interactions under consideration, and the input and post-selected state. In most of the reported works \cite{toppel2013goos,hosten2008observation,goswami2014simultaneous}, the condition $b/a,c/a \ll 1$ allows approximate exponentiation of the response function $F(x,y)$ \cite{toppel2013goos}, eventually making it factorizable of the form (say), $F(x,y)\equiv f(x) g(y)$. As discussed earlier, this factorization of $F(x,y)$ essentially implies the separability of GH and IF shift. However, in certain domains of the mentioned parameters, the response function $F(x,y)$ can not be factorized and eventually the modifications of $x$ and $y$ profiles of the beam, i.e., the GH and IF shifts become non-separable.
\par    
 To observe the non-separability of these two kinds of shifts, we first excite both the effects simultaneously (Eq.\eqref{eq3}) by fixing the input polarization state at $-45^{o}$ (see supplementary information Sec.S.1 for details) \cite{goswami2014simultaneous}. As such modifications are very small in magnitude, we adopt the WVA technique to detect them \cite{hosten2008observation,aharonov1988result,duck1989sense}. The WVA technique is widely associated with the quantification of optical beam shifts \cite{hosten2008observation,goswami2014simultaneous,pal2019experimental}. Note that the protocol for simultaneous amplification of GH and IF shifts were reported previously by Goswami, et al. \cite{goswami2014simultaneous}. However, our objective here is to demonstrate their inherent non-separability. We post-select with the state $\sim[r_p \sin{\epsilon}+ r_s \cos{\epsilon},-r_p \cos{\epsilon}+ r_s\sin{\epsilon}]^T$, where $\epsilon$ is the post-selection parameter describing the overlap between pre and post-selected states \cite{goswami2014simultaneous} (input state is modulated to pre-elected state by the operation of the zeroth order Fresnel matrix  of the interface \cite{gupta2015wave}, see supplementary information Sec.S.1 for details). The spatial response function $F(x,y)$ \cite{asano2016anomalous,pal2019experimental}, here, becomes

\begin{align}
    F(x,y)\sim&(r_p^2+r_s^2)\sin{\epsilon} \Bigg( 1 - \frac{1}{r_p^2+r_s^2} \Bigg[ r_s r_p \cot{\epsilon}\left(\frac{\partial \ln{r_p}}{\partial\theta_i} - \frac{\partial\ln{r_s}}{\partial\theta_i}\right) + \left(r_p^2\frac{\partial\ln{r_p}}{\partial\theta_i} +r_s^2 \frac{\partial\ln{r_s}}{\partial\theta_i}\right) \Bigg] \nonumber \\ &\frac{ix}{z_0+iz} + \frac{r_p+r_s}{r_p^2+r_s^2} \cot{\theta_i} \Bigg[ \cot{\epsilon}(r_p+r_s) + (r_p- r_s)\Bigg]\frac{i y}{z_0+iz}\Bigg)
    \label{eq4}
\end{align}
 Eq.\eqref{eq4} is in the form of Eq.\eqref{eq3}. When pre and post-selection are nearly orthogonal (see supplementary information Sec.S.1 for details) i.e., $\epsilon\rightarrow0$; $\cot{\epsilon}\gg 1$, and $F(x,y)$ can not be factorized. This implies non-separability of the $x$ and $y$ modifications of the post-selected beam, i.e., of the GH and IF shifts respectively.
\begin{figure}[h!]
    \centering
    \includegraphics[width=\textwidth]{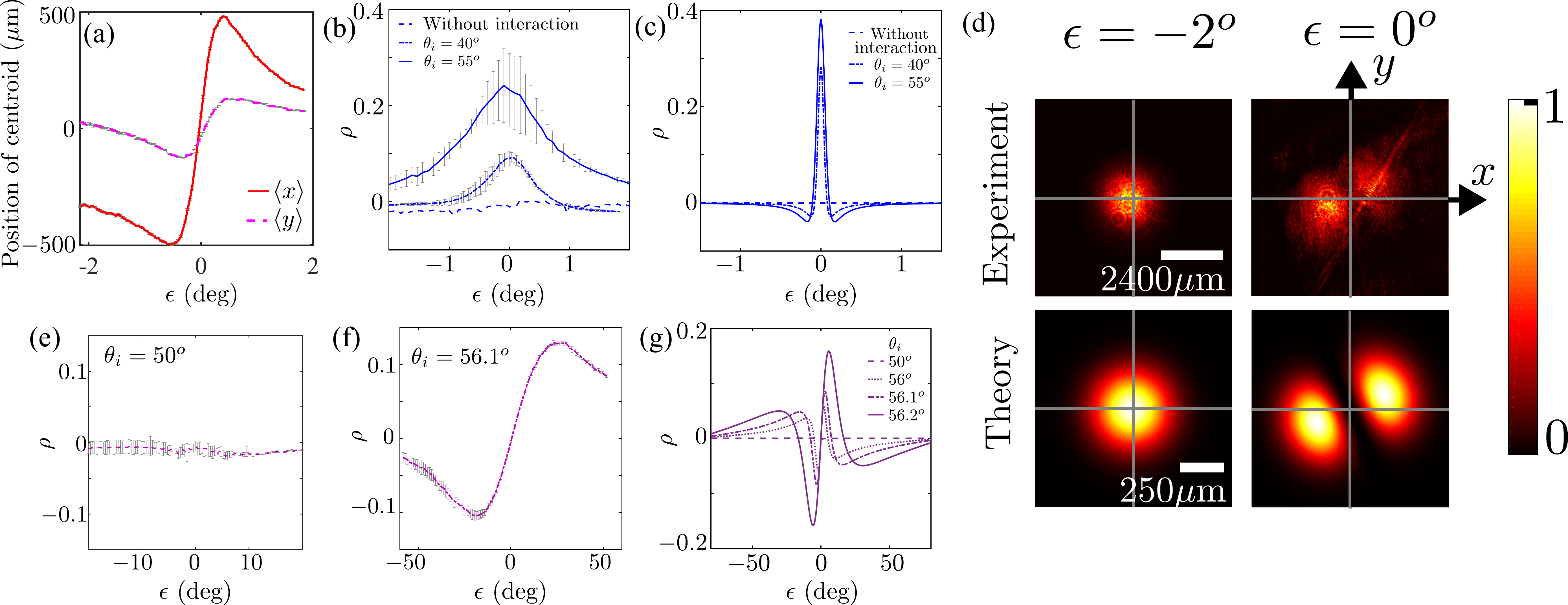}
     \caption{\textbf{Figure. 2: Manifestation of non-separability of the longitudinal (GH) and transverse (IF) shifts of Gaussian beams for $-45^{o}$ ((a)-(d)) and horizontal ((e),(f)) input polarization state.} \textbf{(a)} Variation of the $x$ coordinate ($\langle x \rangle$, red solid line), and $y$ coordinate ($\langle y \rangle$, magenta dashed line) of the centroid with changing post-selection parameter $\epsilon$ at an angle of incidence $\theta_i=40^{o}$. \textbf{(b)} Variation of $\rho$ with changing $\epsilon$ for different $\theta_i=40^{o} \text{ (blue dash-dotted line)},55^{o} \text{ (blue solid line)}$. The variation of $\rho$ with changing $\epsilon$ in absence of any interaction provides the result of control experiment (blue dashed line). \textbf{(c)} Theoretical variation (using Eq.\eqref{eq4}) corresponding to (b). \textbf{(d)} Experimentally and theoretically (using Eq.\eqref{eq4}) obtained beam structure at $\epsilon=-2^{o},0^{o}$ for $\theta_i=55^{o}$ are in agreement. \textbf{(e), (f)} Experimentally obtained variation of $\rho$ with changing $\epsilon$ for $\theta_i=50^{o},56.1^{o}$ respectively. \textbf{(g)} Corresponding theoretical predictions of $\rho$ (using Eq.\eqref{eq5}) for $\theta_i=50^{o} (\text{violet dashed line}), 56^{o}(\text{violet dotted line}), 56.1^{o}(\text{violet dash-dotted line}), 56.2^{o}(\text{violet solid line})$. The error bars represent statistical errors.}
     \label{fig2}
\end{figure}
\par
The non-separability of GH and IF shift for $-45^{o}$ input polarization state is demonstrated in Fig.\ref{fig2}(a)-(d). The $\epsilon$-dependent variations of the $x$ and $y$ coordinates of the centroid of the beam follow standard nature of weak value amplified shift of the beam centroid (Fig.\ref{fig2}(a)) \cite{pal2019experimental,duck1989sense} (the process of extracting $\langle x \rangle$, $\langle y \rangle$, and $\rho$ is discussed in supplementary information Sec.S.2 \cite{thekkadath2016direct}). Both $\langle x \rangle$, and $\langle y \rangle$ are amplified near orthogonal post-selection $\epsilon\rightarrow 0$. Corresponding theoretical plots are given in the supplementary information Sec.S.1. The corresponding non-separability measure $\rho$ ($\approx0$, otherwise) is modulated by the partial reflection from the prism and subsequent post-selection (Fig.\ref{fig2}(b), and (c)). At $\epsilon \rightarrow 0$, the prominent enhancement of $\rho$ is observed as inferred from Eq.\eqref{eq4} (Fig.\ref{fig2}(c)). Also, with changing $\theta_i$, the experimental parameters mentioned in Eq.\eqref{eq4}, varies. Hence, for a particular $\epsilon$, the non-separability also depends on the angle of incidence (Fig.\ref{fig2}(b), and (c)). The beam profile at the largest $\rho$ ($\epsilon = 0^{o}$) is a two-lobe pattern (Fig.\ref{fig2}(d)). The diagonal orientation of such two lobes in the $x-y$ position space points to the existence of position-position classical entanglement \cite{toppel2014classical,puentes2004optical}. Such entanglement occurs as a result of the inherent non-separability of the corresponding DoFs  $x$ and $y$ \cite{forbes2019classically}. This aspect is discussed later. All the experimental results are supported by the corresponding theoretical predictions. In general, experimental variations appear to exhibit slightly broader features as compared to the theoretical predictions. The possible reasons for such deviations are discussed in the supplementary information Sec.S.3.
\par
Now, we consider a scenario where the non-separability of GH and IF shifts appears exclusively due to a change in the angle of incidence $\theta_i$. Accordingly, we take the input polarization state to be horizontal \cite{goswami2014simultaneous}. In such a scenario,  WVA (post-selection parameter $\epsilon\rightarrow 0$) enhances only the IF shift, as the input state is the eigen-polarization of the GH shift operator \cite{goswami2014simultaneous}. The same can also be inferred from the corresponding spatial response function $F(x,y)$ \cite{pal2019experimental}. $F(x,y)$, after the corresponding post-selection \cite{goswami2014simultaneous}, takes the following form.
\begin{equation}
F(x,y)\sim r_p \sin{\epsilon} \left(1 - i\frac{\partial lnr_{p}}{\partial \theta_i}\frac{x}{z_{o}+iz} + i(1+ \frac{r_{s}}{r_{p}})\cot{\theta_i}\cot{\epsilon}\frac{y}{z_{o}+iz} \right)
\label{eq5}
\end{equation}
Although, Eq.\eqref{eq5} suggests modifications in both $x$ and $y$ direction in the reflected beam, the magnitude of the $x$ modification is very small except for a particular interval of $\theta_i$ in the neighbourhood of the Brewster's angle \cite{born2013principles} ($56.31^{o}$ for an air-glass interface). However, the $y$ modifications are relatively much larger at $\epsilon\rightarrow0$ due to WVA ($\cot{\epsilon}$ term in Eq.\eqref{eq5}). Therefore, we can approximate $F(x,y)$ only as a function of $y$ DoF, i.e., $F(x,y)\sim g(y)$; and the separability arises trivially. For $\theta_i=50^{o}$, $\rho\approx0$, irrespective of the value of $\epsilon$ (Fig.\ref{fig2}(e)), indicates separable longitudinal and transverse shifts (centroid shifts are displayed in the supplementary information Sec.S.1). However, when $\theta_i$ approaches the Brewster's angle, $\frac{\partial lnr_{p}}{\partial \theta_i}>1$, which now modifies the $x$ DoF significantly as well. Hence, $F(x,y)$ becomes non-factorizable, and GH and IF shift becomes non-separable. A prominent increase in the non-separability of GH and IF shift (Fig.\ref{fig2}(f)) in such scenario is observed. Here also, the theoretically obtained (using Eq. \eqref{eq5}) results are in qualitative agreement with the experimental observations (Fig.\ref{fig2}(g)). Thus, the \textit{inherent} non-separability of the longitudinal and transverse beam shifts is detected through the non-separability measure $\rho$ by choosing the desired regimes of input polarization, post-selection state and angle of incidence. It is pertinent to emphasize here that the parameter $\rho$ can act as a useful experimental metric for metrological and sensing purposes. The beam shifts are extremely sensitive to the dielectric environment and these have been extensively used for sensing changes in refractive indices, for determining the thickness of thin films, for estimating the number of layers in multi-layered structures and so on  \cite{ling2017recent}. The sensitivity of the beam shifts is the highest near the singular points, such as the Brewster's angle. However, severe deformations in the beam profile near such singular points \cite{aiello2009brewster,merano2009observing} lead to a lack of an absolute reference point. This often poses a problem in quantifying the beam shift parameters, thus limiting the practically achievable sensitivity. In such a situation, the correlation parameter $\rho$ may turn out to be useful in optimizing the achievable sensitivity near these singular regimes, and this therefore holds considerable promise as a novel experimental metric for metrology and sensing (see Supplementary information Sec. S5 for details).  We now turn to demonstrate the direct manifestation of the non-separability of the longitudinal and transverse beam shifts as position-position classical entanglement in the post-selected Gaussian beam \cite{forbes2019classically,puentes2004optical}.

\begin{figure}[h!]
    \centering
    \includegraphics[width=1\textwidth]{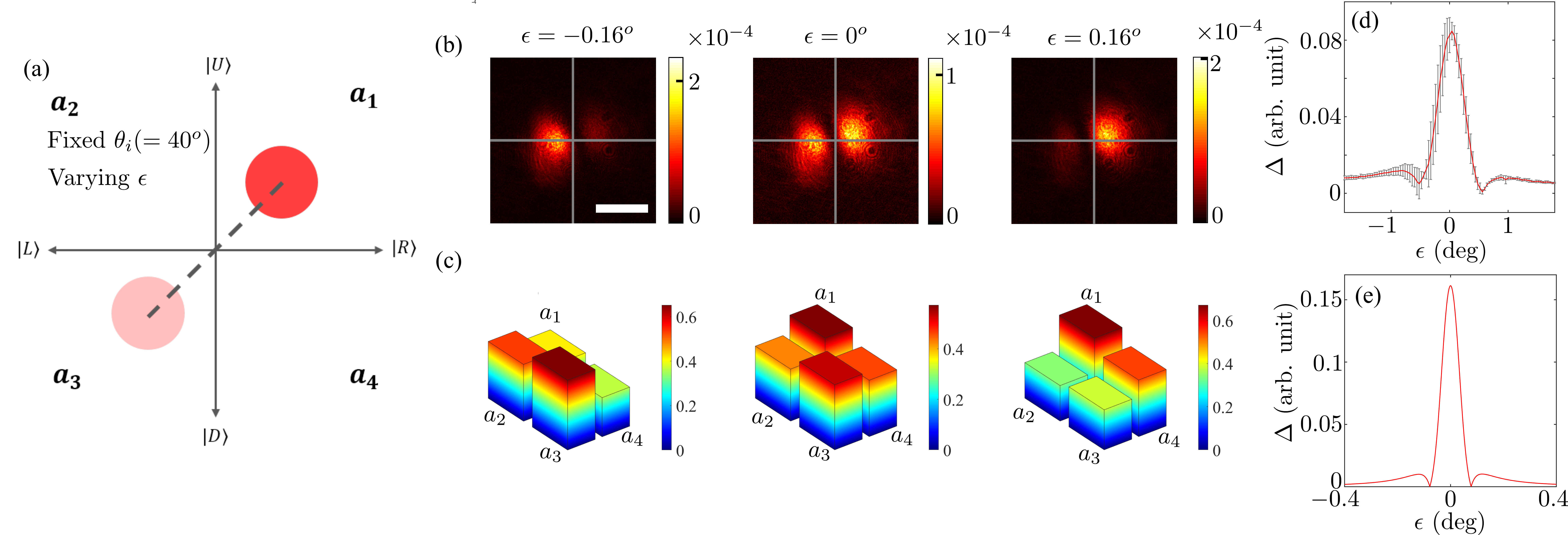}
     \caption{\textbf{Figure. 3: Regulated control of the position-position classical entanglement of partially reflected Gaussian beam by changing the post-selection polarization parameter $\epsilon$ for a fixed angle of incidence $\theta_i=40^{o}$ with input polarization state $-45^{o}$.} \textbf{(a)} Schematic illustration: for a fixed $\theta_i$, changing $\epsilon$ changes the relative intensity of the two intensity lobes. $a_i$s denote the amplitude of the beam in the $i$th quadrant in the $\{R,L,U,D\}$ basis. \textbf{(b)} and \textbf{(c)} Recorded beam structure and the corresponding values of $a_i$ for three different post-selection parameter $\epsilon=-0.16^{o},0^{o}, 0.16^{o}$. At $\epsilon=0^{o}$, the contribution of $a_1,a_3$ become maximal whereas that of $a_2,a_4$ become minimal, demonstrating maximum possible degree of entanglement. White scale bar represents 2400 $\mu m$ length. \textbf{(d)} and \textbf{(e)} Experimentally and theoretically (using Eq.\eqref{eq4}) obtained dependence of $\delta$ with changing $\epsilon$. Maximum possible $\Delta$ appears at $\epsilon=0^{o}$. The error bars represent statistical errors.}
     \label{fig3}
\end{figure}

As mentioned earlier, GH shift is observed in the longitudinal direction (right-left, i.e., $\{R,L\}$) and IF shift occurs in the transverse direction (up-down, i.e., $\{U,D\}$) (see Fig.\ref{fig1}). Any arbitrary beam profile in $x-y$ coordinate system can be interpreted as a two qubit system (first qubit basis: $\{\ket{R},\ket{L}\}$, second qubit basis: $\{\ket{U},\ket{D}\}$) as \cite{toppel2014classical}
\begin{equation}
    \ket{\psi}=a_1\ket{R}\ket{U}+a_2\ket{L}\ket{U}+a_3\ket{L}\ket{D}+a_4\ket{R}\ket{D}
    \label{eq6}
\end{equation}
where $\sum_{i=1}^{4} a_i ^2 = 1$, denotes the normalization of overall state. The values of $a_i$ are the amplitudes of the beam in the $i$th quadrant (see Fig.\ref{fig3}(a)) \cite{toppel2013goos}. $\ket{\psi}$ becomes a maximally entangled state in the following two scenarios, $a_1=1/\sqrt{2}=a_3,a_2=0=a_4;a_1=0=a_3,a_2=1/\sqrt{2}=a_4$ \cite{toppel2013goos}.
As discussed, such entanglement is formally known as position-position entanglement \cite{toppel2013goos}. In the following, we demonstrate such position-position entanglement in our experimental system and control the degree of the entanglement by tuning the experimental parameters. 
We define a quantity 
\begin{equation}
    \Delta=||a_1a_3|-|a_2a_4||
\end{equation} as a measure of the degree of entanglement in the simplistic case of position-position classical entanglement in an optical beam (see supplementary information Sec.S.7 for details). $\Delta=\frac{1}{2}(0)$ indicates maximum (zero) entanglement. Note that $\Delta$ has a correspondence with the previously defined non-separability measure $\rho$. However, $\Delta$ appears to be a more perceptible measure of position-position classical entanglement in a discrete pure state as given in Eq.\eqref{eq6}, estimated just by measuring the intensity of the beam in the four quadrants \cite{toppel2014classical} (see supplementary information Sec.S.4). As mentioned in Fig.\ref{fig2}(d), we get a two-lobe intensity pattern in the post-selected beam when $\epsilon\rightarrow0$. However, these two lobes have a certain intensity distribution, which might spread over all four quadrants. More importantly, for a given angle of incidence $\theta_i$ and pre-selection, the intensity distribution of the two lobes changes as a function of the post-selection parameter $\epsilon$ (see Fig.\ref{fig3}(a)). On the other hand, at a constant $\epsilon=0^{o}$ (say) different $\theta_i$ changes the orientation of the two lobes in the $\{R,L,U,D\}$ basis (see Fig.\ref{fig4}(a), see supplementary information Sec.S.7 for details).  
\begin{figure}[h!]
    \centering
    \includegraphics[width=1\textwidth]{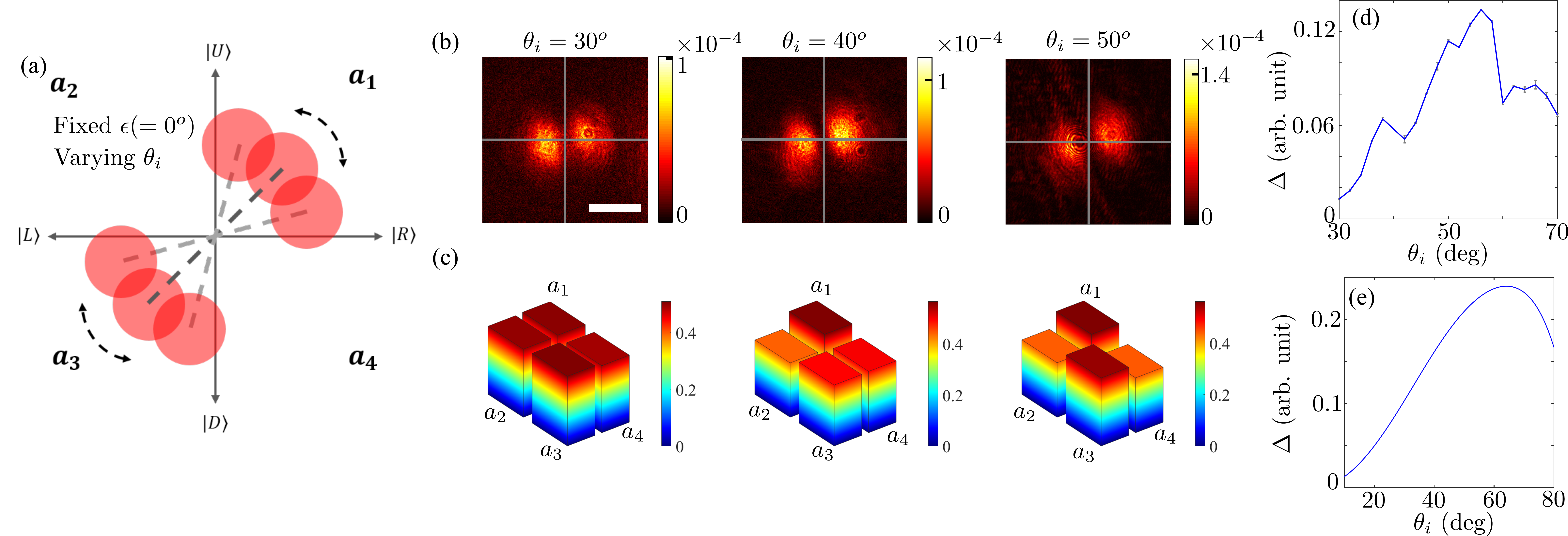}
     \caption{\textbf{Figure. 4: Regulated control of the position-position classical entanglement of partially reflected Gaussian beam by changing the angle of incidence $\theta_{i}$ for post-selection parameter $\epsilon= 0^{o}$ with input polarization state at $-45^{o}$.} \textbf{(a)} Schematic illustration: for  $\epsilon= 0^{o}$, changing $\theta_i$ rotates the intensity distribution of the two-lobe pattern about the defined origin (see supplementary information Sec.S.2). \textbf{(b)} and \textbf{(c)} Recorded beam structure and the corresponding values of $a_i$ for three different angles of incidence $\theta_{i} = 30^{o}$, $\theta_{i} = 40^{o}$ and $\theta_{i} = 50^{o}$. $\theta_{i} = 50^{o}$ shows higher degree of entanglement. White scale bar represents a length of 2400 $\mu m$. \textbf{(d)} and \textbf{(e)} Experimentally and theoretically (Using Eq.\eqref{eq4}) obtained variation of the degree of entanglement $\Delta$ for different angles of incidence $\theta_{i}$. The error bars represent statistical errors.}
     \label{fig4}
\end{figure}
\par
We first demonstrate the dependence of the degree of entanglement on the post-selection parameter $\epsilon$ (Fig.\ref{fig3}). A changing $\epsilon$ changes the relative intensity of the two lobes (Fig\ref{fig3}(b)) and subsequently, the amplitude $a_i$s (see supplementary information Sec.S.4 for the details) of the beam in the four quadrants (Fig.\ref{fig3}(c)). Experimental results and corresponding theoretical agreement (using Eq.\ref{eq4}) suggests that for a fixed $\theta_i$, and input state at $-45^{o}$, the maximum entanglement (in experiment, $\Delta\sim0.09$; in theory, $\Delta\sim0.15$) at $\epsilon=0^{o}$. The possible reasons for the quantitative mismatch between experiment and theory are discussed in supplementary information Sec.S.3.
\par
Next, we demonstrate the dependence of $\Delta$ on the  angle of incidence $\theta_i$ (Fig.\ref{fig4}). By changing $\theta_i$, the corresponding amplitudes $a_1,a_2,a_3,a_4$ in the four quadrants change (Fig.\ref{fig4}(b) and (c)). We observe maximum degree of entanglement $\Delta\sim0.14$ at $\theta_{i} \sim 56^{o}$ (see Fig.\ref{fig4}(d). The corresponding theoretical plot (using Eq.\eqref{eq4}) shows good qualitative agreement with the experimental result (Fig.\ref{fig4}(e)) where $\Delta$ becomes maximum ($\sim0.24$) at $\theta_i\sim 64 ^{o}$. The possible reasons for the mismatch between experiment and theory are discussed in supplementary information Sec.S.3. Therefore, the non-separability of longitudinal and transverse beam shifts paves the way for a two-way control of the degree of the position-position classical entanglement. The control is achieved by tuning the above mentioned experimental parameters, i.e., angle of incidence, and post-selection angle, (as demonstrated in supplementary information Sec.S.7).

\par 
In summary, we have unveiled non-separability of the longitudinal and the transverse optical beam shifts for the simple case of partial reflection of Gaussian beam at dielectric interface. This non-separability, although inherent in the framework of beam shifts, becomes detectable at certain regions in the parameter space, for specific input polarization state, post-selection state, and angle of incidence. The uncovering of the inherent non-separability of these two fundamental optical beam shifts will enrich our understanding on their origin and will also throw new light into a number of related universal wave phenomena \cite{toppel2013goos,asano2016anomalous,bliokh2015spin}. Initial observations indicate that the demonstrated non-separability might be an efficient experimental metric for metrology and sensing \cite{ling2017recent}. It is further demonstrated that this non-separability manifests as position-position classical entanglement in a fundamental Gaussian beam. The degree of entanglement is controlled by tuning the experimental parameters, the post-selection polarization state and the angle of incidence. The demonstrated position-position entanglement along with the polarization state of light opens up an interesting avenue for the generation and controlled manipulation of tripartite classically entangled states using polarized Gaussian beams \cite{forbes2019classically}.

\section*{Acknowledgement}
The authors thank the support of Indian Institute of Science Education and Research Kolkata (IISER-K), Ministry of Education, Government of India. The authors would like to acknowledge the Science and Engineering Research Board (SERB), Government of India, for the funding (grant No. CRG/2019/005558). We like to acknowledge Sayan Ghosh (IISER-K) for scientific discussions that helped to improve our work. We also like to acknowledge the help of Atharva Paranjape for his help in building the experimental setup for the angle of incidence-dependent studies. SG additionally acknowledges  CSIR, Government of India, for research fellowships.

\section*{Conflict of Interest}
The authors declare no conflict of interest.

\bibliographystyle{ieeetr}
\bibliography{ms}

\end{document}


\maketitle
\section{Choice of input and post-selected polarization state}
Real beams carry a finite spectrum of wave vectors around the central wave vector which leads to the shifts in the centroid of its transverse profile when reflected or refracted from any interface \cite{bliokh2013goos}. The longitudinal (to the plane of incidence) Goos-H\"anchen (GH) shift originates from the angular gradient of the Fresnel coefficients associated with the change of angle of incidence for the non-central wave vectors \cite{bliokh2013goos}. The transverse (to the plane of incidence) Imbert-Fedorov (IF) shift originates from the spin orbit interaction arising due to the transformation from the beam coordinate to the global spherical coordinate frame and is associated with the change of the plane of incidence in the local frame \cite{bliokh2013goos}. 
\par
All the above-mentioned transformations of a polarized beam-like electric field can be encapsulated by a momentum domain Jones matrix $M$ \cite{bliokh2013goos}. Note that Eq.(2) of the main text carries the information of $M$ in the electric field of the reflected beam $\Vec{E}_{ref}$. Under certain approximations \cite{toppel2013goos}, this Jones matrix can be written in the following form \cite{aiello2008role,pal2019experimental}.
\begin{equation}\label{eq1}
    M = (\mathbb{I} - i[\frac{x}{z_0+iz} GH + \frac{y}{z_0+iz} IF])T
\end{equation} Where $GH$ and $IF$ represent the shift matrices for GH and IF shift respectively, and $T$ represents the zeroth-order Fresnel reflection Jones matrix \cite{toppel2013goos,born2013principles} of the system under consideration. $GH$ and $IF$ are given by the following $2\cross2$ matrices.

\begin{equation}\label{eq2}
 GH=\left[\begin{array}{cc}
 \frac{\partial \ln r_{p}}{\partial \theta_i} & 0 \\
0 &  \frac{\partial \ln r_{s}}{\partial \theta_i}
\end{array}\right] ,  IF=\left[\begin{array}{cc}
0 & -\left(1+\frac{r_{p}}{r_{s}}\right) \cot \theta_i \\
\left(1+\frac{r_{s}}{r_{p}}\right) \cot \theta_i & 0
\end{array}\right]
\end{equation}
The eigenvalues of the shift matrices provide the maximal centroid shift of the beam when the incident polarization coincides with the corresponding eigenvectors  \cite{toppel2013goos,gotte2014eigenpolarizations}. However, these eigenvalues are extremely small in magnitude for both the shifts in most of the regimes in the corresponding parameter space. As a result, Eq.\eqref{eq1} takes the form of a weak interaction Hamiltonian in which the shift operators $GH$ and $IF$ act on the polarization state of the incident beam \cite{toppel2013goos}. We adopt the technique of weak value amplification (WVA) to enhance these shifts and experimentally detect them \cite{goswami2014simultaneous}. The weak measurement and WVA of the beam shifts is a classical optical analogue of the well-known quantum mechanical post-selected weak measurements \cite{hosten2008observation}. The WVA sequentially involves a pre-selection (served by P1 and subsequent the $T$ matrix of Eq.\eqref{eq1} \cite{gotte2014eigenpolarizations}, see Fig.(1) in the main text), a weak interaction (GH and IF shift from partial reflection from the prism) and post-selection (served by P2). By choosing an appropriate set of pre and post selection before and after the weak measurement process respectively, we can amplify and observe the tiny shifts in the experimental settings. This choice of pre and post-selection is described in the paragraph below.
\par
\begin{figure}[h!]
    \centering
    \includegraphics[width=0.5\textwidth]{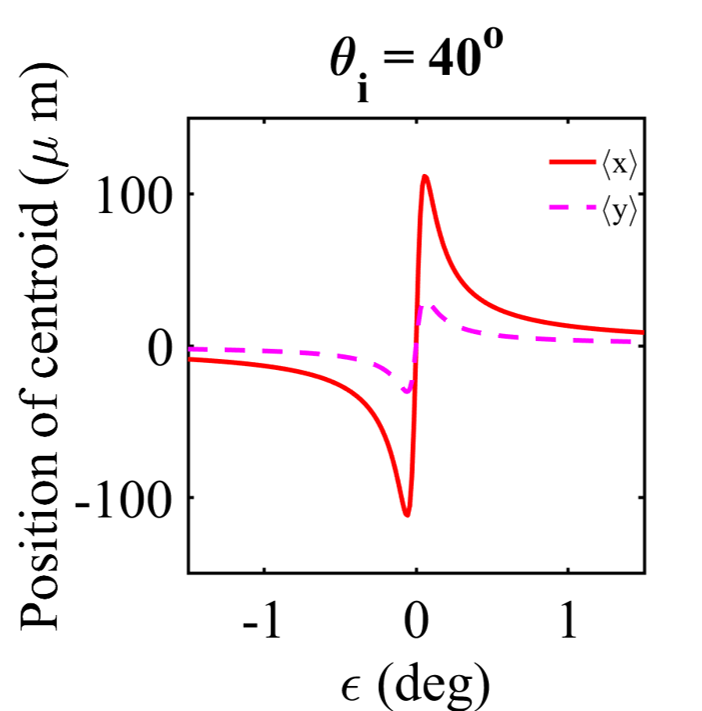}
     \caption{Theoretically obtained variation (using Eq.(4)) of $\langle x \rangle$ (red solid line), $\langle y \rangle$ (magenta dashed line) with changing the post-selection parameter $\epsilon$ at angle of incidence $\theta_i=40^{o}$ and input polarization state at $-45^{o}$.}
     \label{figs1}
\end{figure}

It is quite clear from Eq.\eqref{eq2} that the eigenvectors of the $GH$ matrix are horizontal ($\hat{x}$-polarized) and vertical ($\hat{y}$-polarized) states and that of the $IF$ matrix are linear below the Brewster's angle and elliptical above it \cite{gotte2014eigenpolarizations,goswami2014simultaneous}. In the WVA mechanism, pre-selection is usually chosen as a superposition of both the eigenstates of the measuring observable. We hence intentionally choose the input polarization state as $-45^{o}$ linear polarization to excite both GH and IF shifts \cite{goswami2014simultaneous}. It is important to note that, the input state is modified by the Fresnel reflection matrix, resulting in a pre-selection state given by $\sim[r_{p}, r_{s}]^{T}$. To amplify the shifts, we use a nearly orthogonal post selection given by $\sim[r_{s}\cos{\epsilon}+\ r_{p}\sin{\epsilon}, -r_{p}\cos{\epsilon} +\ r_{s}\sin{\epsilon}]^{T}$. Using Eq.(4) of the main text, we calculate (the value of the corresponding parameters are mentioned in Sec.S.3) the $x$ and $y$ coordinate of the position of the centroid ($\langle x\rangle, and \langle y\rangle$) with changing the post-selection parameter $\epsilon$. Fig.\ref{figs1} demonstrates the variation of $\langle x\rangle, \langle y\rangle$ with changing $\epsilon$ for angle of incidence $\theta_i=40^{o}$.
\par
\begin{figure}[h!]
    \centering
    \includegraphics[width=0.5\textwidth]{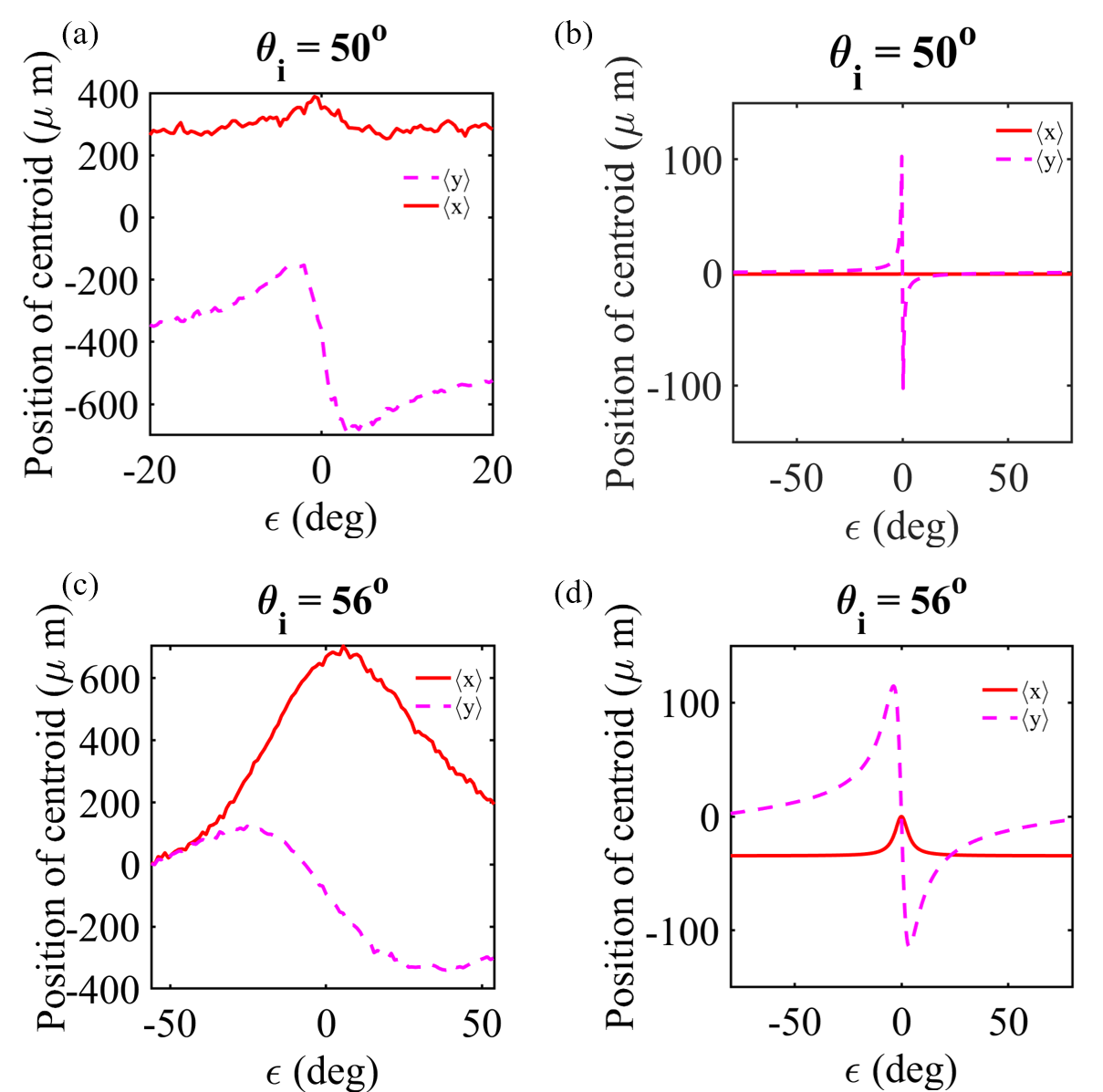}
     \caption{\textbf{Theoretically and experimentally obtained variations of the position of centroid for horizontal input polarization state} \textbf{(a)} Experimentally obtained variation of the $x$ coordinate ($\langle x \rangle$, red solid line), and $y$ coordinate ($\langle y \rangle$, magenta dashed line) of the centroid with changing post-selection parameter $\epsilon$ at an angle of incidence $\theta_i=50^{o}$. \textbf{(b)} Corresponding theoretical variation (using Eq.(5) of the main text). \textbf{(c)} Experimentally obtained variation of the  ($\langle x \rangle$, red solid line), and ($\langle y \rangle$, magenta dashed line) of the centroid with changing post-selection parameter $\epsilon$ at an angle of incidence $\theta_i=56^{o}$. \textbf{(d)} Corresponding theoretical variation (using Eq.(5) of the main text).}
     \label{figs2}
\end{figure}

 
Next, to selectively excite the IF shift, we consider the input polarization horizontal \cite{goswami2014simultaneous}. With the corresponding near orthogonal post-selection $\sim[-\sin{\epsilon},\cos{\epsilon}]^T$,
 the theoretically (using Eq.(5) of the main text) and experimentally obtained variation of the $x$ coordinate ($\langle x \rangle$), and $y$ coordinate ($\langle y \rangle$) of the centroid with changing post-selection parameter $\epsilon$ are plotted in Fig.\ref{figs2} for angle of incidence $\theta_i=50^{o},56^{o}$. It is evident that when $\theta_i$ approaches Brewster's angle ($\sim 56.31^{o}$ for an air-glass interface) $\langle x\rangle$ increases rapidly and the non-separability of longitudinal and transverse beam shift becomes prominent (see Fig.2(f) and (g) of the main text). The possible origins of the mismatches between experimental and theoretical results are discussed in Sec.S.3. 
 \par
 Note that, all the theoretical calculations are carried out using the general form of the reflected vector Gaussian beam (Eq.(2) of the main text) following \cite{aiello2008role}. However, we invert the $x$ coordinate of all the theoretical outputs to fit it with our experimental convention of coordinates (Fig.1 of the main text).

\section{Image analysis}\label{img}

As mentioned in the main text, we analyse the image to obtain the desired quantities. We use MATLAB to analyze the image. The chip size of the camera used is 512$\cross$512. Hence, the image is read as a 512$\cross$512 matrix in MATLAB where each element corresponds to a pixel of the image and the value of that element corresponds to the intensity of that pixel. In the following, we discuss the procedure to estimate $\langle x\rangle, \langle y\rangle, \rho$ from experimentally obtained image of the light beam.

\begin{enumerate}

\item Background subtraction:

To nullify the contribution of stray lights (noise) in the detected image, we take the background signal blocking the reflected beam. This image is taken at the beginning of each experiment, and we do not change the experimental setting once this image is taken. We then element wise subtract the intensity of the reference beam from that of a recorded image so that the image taken for analysis doesn't have any signature of any constant source of background noise.

\item Coordinate transformation:

 In our experiment, we take the centroid of the unshifted beam ($x_0,y_0$) (centroid of the image taken while the pre and post-selection are same) as the reference i.e. the centroid of this beam is set as origin of a Cartesian coordinate system. We transform the coordinate system of the CCD sensor to the centroid of the unshifted image to define our working Cartesian system. Then, we represent any given image in that coordinate system. This process is executed by modifying $x$ and $y$ of any image with $x-x_0$ and $y-y_0$ respectively for $x$ and $y$ coordinate of each pixel. Thus, we obtain the actual value of $\langle x \rangle$, $\langle y \rangle$, and $\rho$ with respect to the unshifted beam.

\item Cropping: 

We crop all the images before starting the analysis with reference to the centroid of the unshifted beam ($x_0$, $y_0$) (centroid of the image).

\item Finding the centroids:

To find the centroid, we carry out an element-wise discrete sum of the coordinates ($x$ or $y$) over the intensity profile of the image. The centroid in the $x$ and $y$ direction and their correlation can be defined respectively as follows
\begin{equation}\label{cen}
   \langle x \rangle = \frac{\sum (x-x_0)I }{\sum I},  \langle y \rangle = \frac{\sum (y-y_0)I }{\sum I}
\end{equation}
\begin{equation}\label{xy}
    \langle xy \rangle = \frac{\sum (x-x_0)(y-y_0)I }{\sum I}
\end{equation}
where I is the intensity of the corresponding image. The position $x-x_0$, and $y-y_0$ is initially calculated in unit of pixel numbers, each of which is 24 $\mu m$ in size. So, to extract the aforementioned quantities in a physically realizable unit (say, in SI unit), we need to multiply $x-x_0$, and $y-y_0$ by $24 \cross 10^{-6}$.

\item Defining the measure of correlation :

To study the non-separability between the $x$ and $y$ degrees of freedom of the beam, we use $\rho$ (see Eq.(1) of the main text). From the experimental images, $\rho$ can be calculated as follows.
\begin{equation}\label{rho}
    \rho = \frac{\langle xy \rangle -\langle x \rangle\langle y \rangle}{\sqrt{\langle x^2 \rangle  - \langle x \rangle^2}\sqrt{\langle y^2 \rangle  - \langle y \rangle^2}}
\end{equation}

\end{enumerate}
In theoretical calculation, the same process is followed and $\rho$ is calculated in the same way from the simulated post-selected beam (using Eq.(4), (5) of the main text with the parameters mentioned in Sec.S.3). A similar process of image analysis to extract $\langle xy \rangle$ was previously demonstrated in \cite{thekkadath2016direct}. However, the objective of that work was different.
\section{Experimental parameters and discussions on the mismatch between theoretical and experimental results}
The experimental parameters are $z=14\ cm$, $z_0=7\ cm$, and refractive index of the prism is taken $1.5$. The Rayleigh range $z_0$ is calculated from beam waist at focus of the lens L (see Fig.1 of the main text). The beam waist is estimated from the image recorded by putting the camera at the focal plane of L. These values of the parameters were included during the simulation of Eq.(4) and (5) of the main text.
\begin{figure}[h!]
    \centering
    \includegraphics[width=1\textwidth]{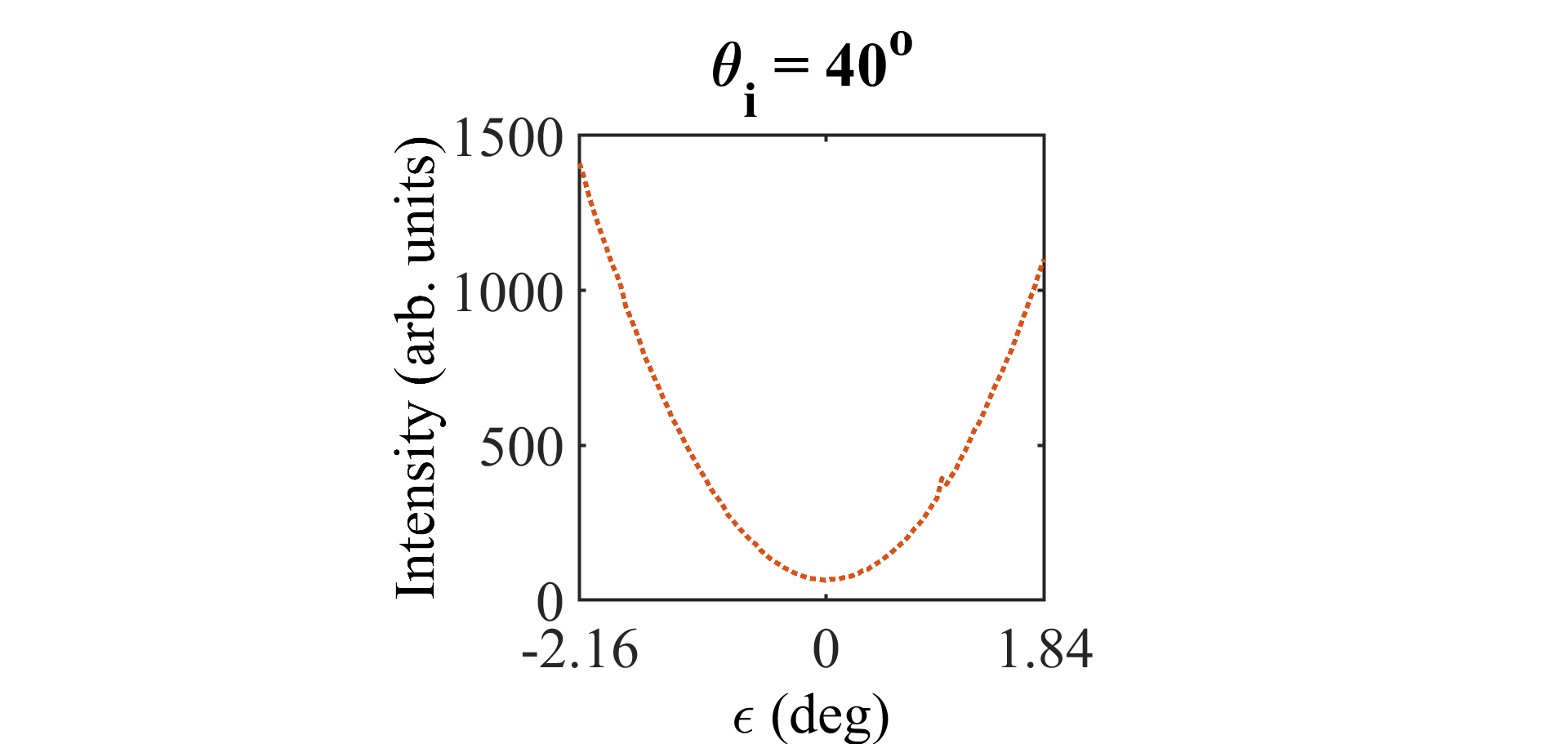}
     \caption{Variation of the intensity of the post-selected beam with changing post-selection parameter $\epsilon$ at an angle of incidence $\theta_i=40^{o}$ and $-45^{o}$ input polarization state.}
     \label{figs3}
\end{figure}

\par
$z_0$ is estimated measuring the beam waist at focus \cite{born2013principles}. To measure that we put the camera at the focus of the lens L. However, the beam waist at focus becomes very narrow ($\sim170\mu m$). Hence, this estimation is erroneous due to the limitation in the spatial resolution of the CCD camera used ($\sim24\mu m$). Also, errors might appear in calculating $z$ as the detector chip is installed at a certain distance from the face, inside the cabinet of the camera. 
\par 
Another unavoidable error contributor is the noise appearing due to the low level of detected intensity. WVA is performed at near orthogonal pre and post-selection configuration which reduces the intensity of the post-selected beam significantly \cite{aharonov1988result,duck1989sense,goswami2014simultaneous,hosten2008observation}. For $-45^{o}$ input polarization, at $\theta_i=40^{o}$, the variation of the detected intensity is plotted with changing $\epsilon$ which shows certain decrease in the intensity level around near orthogonal post-selection region $\epsilon\rightarrow0$ (Fig. \ref{figs3}). On the other hand, for the horizontal input state, when $\theta_i$ approaches the Brewster's angle, the air-glass interface itself reduces the intensity of the reflected beam rapidly \cite{gotte2014eigenpolarizations}. Again, the near orthogonal post-selection makes the beam more dim. These low levels of intensities usually affect the estimation of any parameter from the intensity of the beam, such as, $\langle x \rangle, \langle y \rangle$ etc. It is also to be noted that around Brewster's angle and subsequent near orthogonal post-selection, the beam structure gets distorted, which causes errors in calculating the centroids \cite{aiello2009brewster}.
\par
Another possible reason for the deviation in experimental results from the theoretical predictions are the inbuilt assumptions in Eq.(2) of the main text. As discussed in the main text, the GH and IF shifts, related to the $x$ and $y$ modification of the reflected beam respectively, are formally assumed to to have completely distinct origin \cite{bliokh2013goos}. This assumption provides considerably good agreement with the experiment at most of the regions of the corresponding parameter space. However, if the exact treatment is adopted (see the supplemental material of \cite{zhu2021wave}), the theoretical predictions are expected to be more realistic. That said, we note that the approximated equation (Eq.(2)) is sufficient for the understanding the non-separability and subsequent position-position classical entanglement presented in the paper.

\section{Calculation of $a_1,a_2,a_3,a_4$ }

As discussed in the main text, longitudinal $\{\ket{R}, \ket{L}\}$ and transverse $\{\ket{U}, \ket{D}\}$ direction are used as two qubits of the bipartite system. The strength of entanglement depends upon the amplitude $a_1$, $a_2$, $a_3$ and $a_4$ of the four possible composite states $\ket{RU}$, $\ket{LU}$, $\ket{LD}$ and $\ket{RD}$ respectively. These amplitudes are obtained from the intensities in the corresponding four quadrants of the image recorded in the experiment. These quadrants are of the Cartesian coordinate defined by the centroid of the unshifted beam ($x_0,y_0$) as the origin. First, the image is area-normalized by standard procedure to ensure the normalization of $\ket{\psi_f}$, i.e. $\sum_{i=1}^{4} I_i = 1$, where $I_i$ is the intensity of the $i$th quadrant of the unnormalized image. Once normalization is done, the square root of intensity of $i$th quadrant provides $a_i$. In theoretical calculation, $a_1,a_2,a_3,a_4$
are calculated in the same way from the simulated post-selected beam (using Eq.(4) of the main text with the parameters mentioned in Sec.S.3).

\section{Estimation of sensitivity of $\rho$}
\begin{figure}[h!]
    \centering
    \includegraphics[width=1\textwidth]{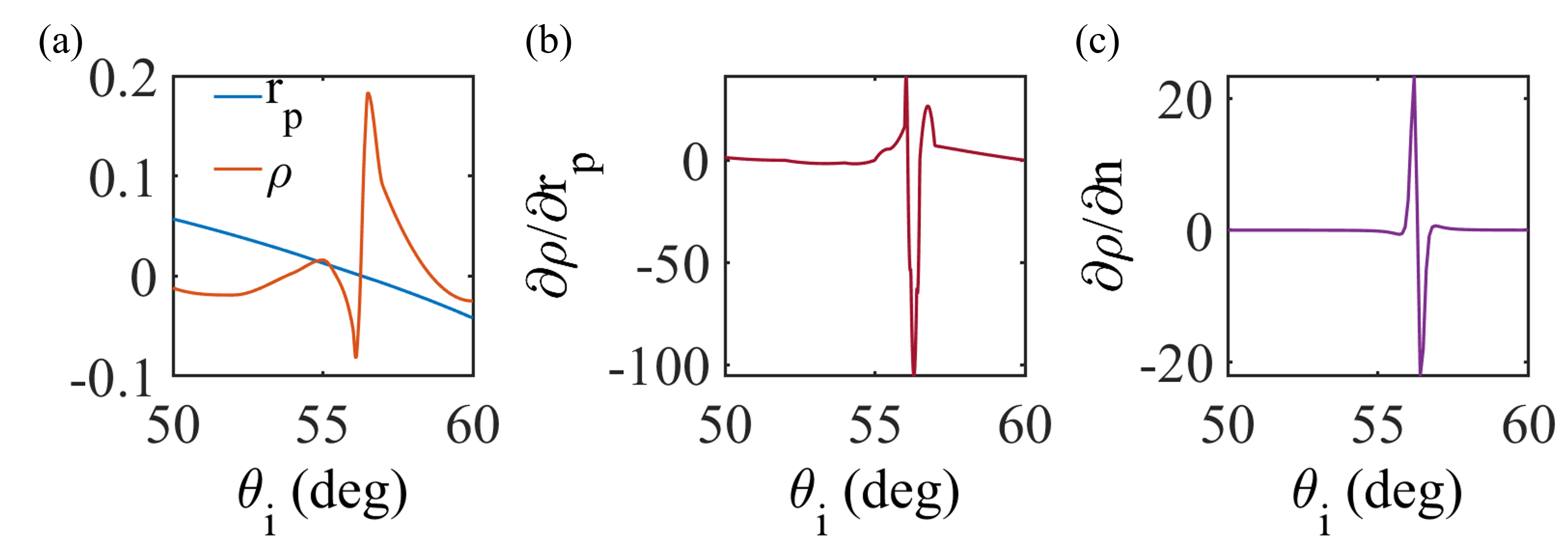}
     \caption{\textbf{Demonstration of $\rho$ as a sensitive experimental metric.} (a) Experimentally obtained variation of $\rho$ (orange solid line) with angle of incidence $\theta_i$ for a fixed input and post-selection state. Intermediate data points are interpolated. Calculated \cite{gupta2015wave} variation of Fresnel reflection coefficient $r_p$ (blue solid line) with angle of incidence $\theta_i$. (b) Corresponding variation of $\frac{\partial\rho}{\partial r_p}$ with changing $\theta_i$ indicates the enhancement of the sensitivity of $\rho$ towards varying $r_p$ in the proximity of the Brewster's angle. (c) Theoretically obtained (using Eq.(5) of the main text) variation of $\frac{\partial\rho}{\partial n}$  with $\theta_i$.} 
     \label{figs5}
\end{figure}

GH and IF shifts have shown promise in various metrological applications, such as measuring the thickness of metallic films, the number of Graphene layers in a multi-layer structure, etc \cite{ling2017recent}. The sensitivity of these shifts towards change in refractive index $n$ attains its maximum around the singular points such as, Brewster's angle in the case of partial reflection. However, the beam profile gets severely deformed  \cite{aiello2009brewster,merano2009observing} and hence, there is a lack of an absolute reference point near such points. This makes the estimation of the shifts erroneous and sets a practical limit on the sensitivity of the shifts towards the change in refractive index $n$. In contrast, $\rho$ depends on the beam profile and can be obtained by exploiting both the variants of the shifts without resorting to their explicit quantification. As shown in (Fig.\ref{figs5}(a), (b)), around Brewster's angle of incidence, $\rho$ becomes significantly sensitive towards even a minute change in the Fresnel reflection coefficient $r_p$. Note that as $r_p$ is directly related to refractive index $n$, the presented experimental results indicate the increase in the corresponding sensitivity of $\rho$ towards $n$. This is corroborated by the theoretical variation of sensitivity of $\rho$ towards changing $n$, $\frac{\partial \rho}{\partial n}$ (Fig.\ref{figs5}(c)). Therefore $\rho$ can act as a useful experimental metric in metrology, specifically around the singular region in the parameter space. An extensive experimental study on the refractive index sensitivity of $\rho$ is currently underway.

\section{Quantification of degree of entanglement}

 We have demonstrated $\Delta$ as a simplistic measure of the degree of entanglement. $\Delta$ has one to one correlation with Schmidt number \cite{toppel2014classical}, i.e., for maximum degree of entanglement, $\Delta=0.5$, Schmidt number $=2$; $\Delta=0$, Schmidt number $=1$ indicates absence of any entanglement. We note that $\Delta$ efficiently serves the role of the entanglement measure in our case of position-position classical entanglement. However, $\Delta$ might not universally qualify as a measure of the degree of entanglement, specifically, if the feature of non-locality is included.

\section{Dependence of position-position classical entanglement on different experimental parameters}
It is apparent that the strength of the position-position entanglement depends upon the parameters $a_1,a_2, a_3, a_4$ which are determined by the intensities in the four quadrants of the $\{R,L,U,D\}$ coordinate system. These intensities depend on the relative simultaneous contributions of GH and IF shifts which, again, depends upon several experimental parameters, such as, incident angle $\theta_i$, pre-selection polarization, and post-selection parameter $\epsilon$. Here we briefly discuss the contributions of these experimental parameters towards controlling the entanglement strength. 
\par
A changing incident angle changes the experimental parameters $r_p, r_s, \frac{\partial r_p}{\partial \theta_i}, \frac{\partial r_p}{\partial \theta_i}$ under consideration. The amplitudes of GH and IF shifts are strongly dependent on these parameters. Moreover, IF shift also depends upon the incident angle $\theta_i$ itself. Thus, a changing incident angle effects the GH and IF shifts both by means of modifying the matrix elements of $M$ and the eigenvalues and eigenvectors of the corresponding shift matrices (see Eq.(2)) \cite{bliokh2013goos,toppel2013goos}.
\begin{figure}[h!]
    \centering
    \includegraphics[width=0.5\textwidth]{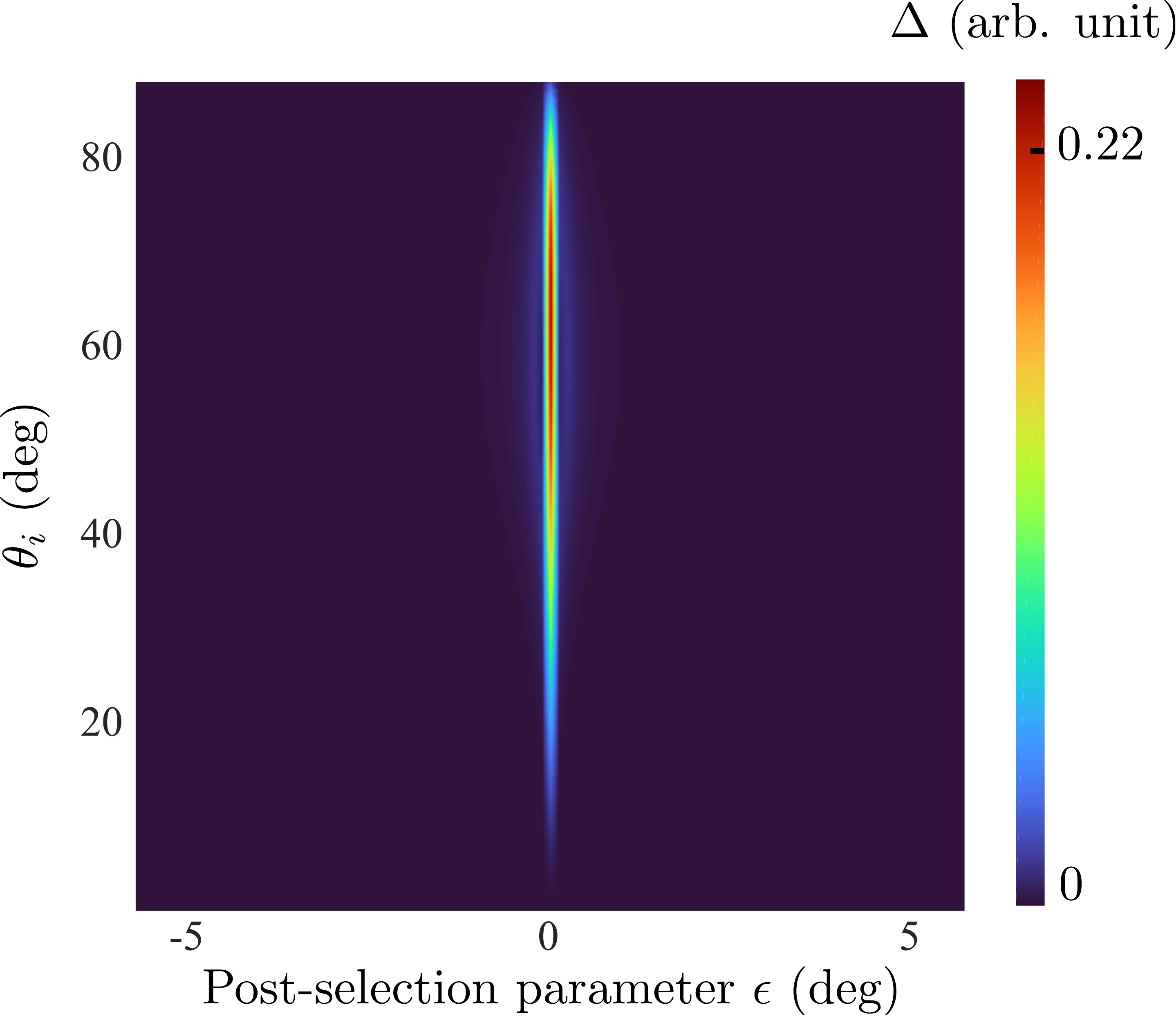}
     \caption{\textbf{Two-way control of position-position entanglement in a partially reflected Gaussian beam}. Theoretically obtained (using Eq.(4) of  main text) dependence of the degree of entanglement $\Delta$ on the angle of incidence $\theta_i$ and post-selection parameter $\epsilon$ for $-45^{o}$ input polarization state. Maximum entanglement is achieved at $\epsilon=0^{o},\text{and}\ \theta_i\sim60^{o}$.}
     \label{figs4}
\end{figure}
\par
For a given $\theta_i$ and input polarization, the strength of the position-position entanglement changes with varying post-selection parameter $\epsilon$. Maximum entanglement occurs when the two intensity lobes become equally intense. Such a scenario is observed at $\epsilon=0$ for any weak value amplified beam shift detection \cite{duck1989sense}. 
\par
Thus, tuning these two parameters, one can control the strength of position-position entanglement (Fig.\ref{figs4}).

\bibliographystyle{ieeetr}
\bibliography{supplement}